\documentstyle[11pt,epsfig]{article}
\begin{document}
\title{Charge density distributions and related form factors in
neutron-rich light exotic nuclei}
\author{A.N. Antonov$^{a,b}$, M.K. Gaidarov$^{a}$, D.N. Kadrev$^a$,\\
P.E. Hodgson$^c$, E. Moya de Guerra$^d$}
\date{$^a${\it Institute of Nuclear Research and Nuclear Energy,\\
         Bulgarian Academy of Sciences, Sofia 1784, Bulgaria}\\
$^b${\it Departamento de Fisica Atomica, Molecular y Nuclear,\\
Facultad de Ciencias  Fisicas, Universidad Complutense de
Madrid,\\
Madrid E-28040, Spain}\\
$^c${\it Subdepartment of Particle Physics, Department of Physics,\\
University of Oxford, Oxford OX1-3RH, United Kingdom}\\
$^d${\it Instituto de Estructura de la Materia, CSIC, Serrano
123,\\
28006 Madrid, Spain}}
\maketitle

Charge form factors corresponding to proton density distributions
in exotic nuclei, such as $^{6,8}$He, $^{11}$Li, $^{17,19}$B and
$^{14}$Be are calculated and compared. The results can be used as
tests of various theoretical models for the exotic nuclei
structure in possible future experiments using a colliding
electron-exotic nucleus storage ring. The result of such a
comparison would show the effect of the neutron halo or skin on
the proton distributions in exotic nuclei.

\section{Introduction}
It has been shown (see, e.g. the review \cite{Cas2000}) that the
study of exotic nuclei is likely to reveal new aspects of nuclear
structure. Since the first experiments with radioactive nuclear
beams (RNB) took place (e.g. \cite{Tani85}) it has been found from
analyses of the total interaction cross sections that weakly-bound
nuclei have increased sizes that deviate substantially from the
$R\sim A^{-1/3}$ rule. It became clear (e.g.
\cite{Hansen95,Doba94}) that such new phenomena are due to the
weak binding of the last few nucleons which form a diffuse nuclear
cloud due to quantum-mechanical penetration (the so called nuclear
halo). Another effect is that the nucleons can form a neutron skin
\cite{Tani95} when the neutrons are on average less bound than the
protons. Most of the neutron haloes have been studied for the
lightest elements, such as He, Li, Be, B and other isotopes.

Direct reactions are a well-known tool to provide information on
nuclear structure. Most exotic nuclei are so shortlived that they
cannot be used as targets. Instead, direct reactions with RNB can
be done in inverse kinematics where the role of beam and target
are interchanged. For example, the proton elastic-scattering
angular distributions were measured for several incident energies,
namely 25.1$A$ MeV \cite{Ogan199,Ogan299}, 41.6$A$ MeV
\cite{Cort97,Lago2001,Cort96} and 71$A$ MeV
\cite{Korsh197,Korsh297} for $^{6}$He, 26.25$A$ MeV (the
preliminary data obtained in Dubna \cite{Ter2001}), 32$A$ MeV
\cite{Korsh197,Korsh297}, 66$A$ MeV \cite{Korsh197,Korsh297} and
73$A$ MeV \cite{Korsh197,Korsh297,Chul95,Korsh93} for $^{8}$He, as
well as for energies less than 100$A$ MeV \cite{Korsh197,Cas2000}
in the cases of $^{6}$Li, $^{11}$Li and others. In some of the
theoretical analyses (e.g.
\cite{Korsh197,Korsh297,Chul95,Crespo95}) the eikonal approach
using neutron and proton density distributions as well as
parametrized nucleon-nucleon total cross sections have been used.
The real part of the optical potential for calculations of
$^{6}$He+$p$, $^{6}$He+$^{4}$He ($E_{lab}$=151 MeV)
\cite{Avrig2000} and $^{6}$He+$p$, $^{8}$He+$p$ ($E_{inc}<100$$A$
MeV) \cite{Avrig2002} elastic scattering angular distributions was
obtained microscopically using the realistic M3Y-Paris effective
interaction \cite{Anant83,Sat83,Dao2000} as well as the Tanihata
neutron and proton densities of helium isotopes \cite{Tani92} in
\cite{Avrig2000,Avrig2002} and the densities derived from the
cluster-orbital shell model approximation (COSMA)
\cite{Zhuk93,Korsh197,Korsh297,Zhuk94} in \cite{Avrig2002}. It was
shown that elastic scattering is a good tool to distinguish
between different density distributions \cite{Avrig2002}. The
results of \cite{Avrig2002} were compared also with those from the
alpha-core approach with the complex and fully non-local effective
interaction \cite{Dort98} and the no-core model based on the large
space shell-model calculations
(LSSM)(\cite{Kara2000,Lago2001,Amos2000} and refs. therein). For
larger energies($>500$ MeV/nucleon) it is accepted that the
Glauber approximation is a suitable method to study charge and
matter distributions from proton elastic scattering data
\cite{Cas2000,Alkha97}.

As mentioned in \cite{Cas2000}, however, an accurate determination
of charge distributions of exotic nuclei can be obtained from
electron-nucleus scattering in inverse kinematics using a
colliding electron-exotic nucleus storage ring.

The aim of the present paper is to calculate the charge form
factors of exotic nuclei such as $^{6}$He, $^{8}$He, $^{11}$Li,
$^{17}$B, $^{19}$B and $^{14}$Be on the basis of charge density
distributions calculated by using various theoretical models. This
work can serve as a challenge for future experimental works on
measurements of the charge form factors and thus for accurate
determination of the charge distributions in these nuclei. The
latter can be a test of the different theoretical models used for
obtaining charge distributions. Second, this work can throw light
on the question about the effect of the neutron halo or skin on
the proton distributions in the helium, litium, boron and
beryllium isotopes.

A brief representation of the theoretical scheme is given in Sec.
II. The results and discussion are presented in Sec. III.

\section{The theoretical scheme}
The charge form factor $F_{ch}({\bf q})$ is the Fourier transform
of the charge density distribution $\rho_{ch}({\bf r})$:
\begin{equation}
F_{ch}({\bf q})=\int d{\bf r} \rho_{ch}({\bf r}) e^{i{\bf qr}}
\frac{1}{\int \rho_{ch}({\bf r}) d{\bf r}}.
\label{eq:ffdef}
\end{equation}
Let us firstly consider $\rho_{ch}({\bf r})$ obtained from a total
nuclear wave function in which the positions ${\bf r}$ of the
nucleons are defined with respect to the centre-of-mass of the
nucleons in the nucleus (c.m.). If the charge density is
spherically symmetrical and normalized to unity, then:
\begin{equation}
F_{ch}({\bf q})=4\pi \int \rho_{ch}(r) j_{0}(qr) r^{2} dr,
\label{eq:ffspher}
\end{equation}
where $j_{0}(qr)$ is the zero-order spherical Bessel function.
Neglecting the charge neutron form factor, the charge density
$\rho_{ch}({\bf r})$ is represented by the density distribution of
point-like protons in the nucleus $\rho_{point}({\bf r})$ folded
with the charge distribution of the proton $\rho_{pr}({\bf r})$:
\begin{equation}
\rho_{ch}({\bf r})=\int \rho_{point}({\bf r}-{\bf r}^{\prime})
\rho_{pr}({\bf r}^{\prime}) d{\bf r}^{\prime}.
\label{eq:rhofolded}
\end{equation}
Then the charge form factor (\ref{eq:ffspher}) can be obtained in
the form:
\begin{equation}
F_{ch}({\bf q})=F_{point}({\bf q}) F_{pr}({\bf q}),
\label{eq:fffolded}
\end{equation}
where $F_{point}({\bf q})$ and $F_{pr}({\bf q})$ are the form
factors which correspond to the densities $\rho_{point}({\bf r})$
and $\rho_{pr}({\bf r})$ (in formulae such as Eq.
(\ref{eq:ffspher})), respectively. Secondly, in some calculations
the wave functions which lead to the density distributions are
often translationally non-invariant and the positions of the
nucleons are determined with respect to the centre of the
potential related to the laboratory system (lab.) but not to the
centre-of-mass of the nucleons in the nucleus. These wave
functions admit a motion of the centre-of-mass which has to be
switched off in the calculations. It is shown (e.g. in
\cite{Forest66,Alkha91,Burov77}) that the relationship between the
form factor calculated by means of harmonic-oscillator wave
functions (with all the nucleons in the $1s$ state)
$F_{point,lab}({\bf q})$ and the form factor calculated by actual
wave functions (with positions of the nucleons determined in the
c.m. system) $F_{point,c.m.}({\bf q})$ contains a correction
factor:
\begin{equation}
F_{point,c.m.}({\bf q})=e^{(qb)^{2}/4A} F_{point,lab}({\bf q}),
\label{eq:correctho}
\end{equation}
where $b$ is the harmonic-oscillator parameter, or
\begin{equation}
F_{point,c.m.}({\bf q})=e^{(qR)^{2}/6A} F_{point,lab}({\bf q}),
\label{eq:correctrms}
\end{equation}
$R$ being the root-mean-square (rms) radius of the nucleus. For
shell-model potentials different from the harmonic-oscillator one
the Eqs. (\ref{eq:correctho}) and (\ref{eq:correctrms}) are
approximate. Thus in this case the expression for the form factor
which can be compared with the experimental form factor can be
written in the form \cite{Burov77}:
\begin{equation}
F_{ch}({\bf q})=e^{(qR)^{2}/6A} F_{point,lab}({\bf q}) F_{pr}({\bf
q}). \label{eq:ffexp}
\end{equation}
In the Eqs. (\ref{eq:ffdef}), (\ref{eq:ffspher}),
(\ref{eq:fffolded}) and (\ref{eq:ffexp}) all form factors are
equal to unity at ${\bf q}$=0.

\section{Results and discussion}
In this Section we calculate $F_{point}({\bf q})$ using the point
charge density $\rho_{point}({\bf r})$ in Eq. (\ref{eq:ffspher})
and $F_{ch}({\bf q})$ using Eq. (\ref{eq:fffolded}) for the exotic
nuclei $^{6,8}$He, $^{11}$Li, $^{14}$Be and $^{17,19}$B. For
comparison we present also the charge form factors of the
"non-exotic" $^{4}$He and $^{6}$Li nuclei.

Various charge distributions of the proton $\rho_{pr}({\bf r})$
can be used in order to obtain the form factor $F_{pr}({\bf q})$.
We use the Gaussian form normalized to unity:
\begin{equation}
\rho_{pr}(r)=\frac{1}{(\pi
\alpha^{2})^{3/2}}\exp(-r^{2}/\alpha^{2})
\label{eq:gaussrho}
\end{equation}
which leads to the proton form factor
\begin{equation}
F_{pr}({\bf q})=e^{-q^{2}\alpha^{2}/4}.
\label{eq:ffgauss}
\end{equation}
For the value of the parameter $\alpha$ we choose $\alpha=0.6532$
fm which corresponds to the proton rms radius $R_{p}=0.80$ fm
\cite{Horn75}.

\subsection{Density distributions of $^{4,6,8}$He}

We use firstly the following point nucleon densities
$\rho_{point}(r)$ for $^{4,6,8}$He deduced by Tanihata et al.
\cite{Tani92} which determine $F_{point}({\bf q})$ in Eq.
(\ref{eq:fffolded}):
\begin{equation}
\rho_{point}^{X}(r)=\frac{2}{\pi^{3/2}}\left \{\frac{1}{a^{3}}\exp
\left [-\left(\frac{r}{a}\right )^{2} \right
]+\frac{1}{b^{3}}\frac{X-2}{3}\left(\frac{r}{b}\right)^{2}\exp
\left [-\left(\frac{r}{b}\right )^{2} \right ]\right \},
\label{eq:rhotani}
\end{equation}
where $X=Z,N$. These densities were obtained by fitting the
measured total reaction cross sections of $^{4,6,8}$He at 800$A$
MeV incident on the C target using the optical limit of the
Glauber model \cite{Karol75}. The parameter values $a$ and $b$ are
determined from:
\begin{equation}
a^{2}=a^{*2}\left(1-\frac{1}{A}\right ), \;\;\;\;\;\;
b^{2}=b^{*2}\left(1-\frac{1}{A}\right ),
\label{eq:glauber}
\end{equation}
where $a^{*}=1.53$ fm for all $^{4,6,8}$He isotopes, $b^{*}=2.24$
fm for $^{6}$He and $b^{*}=2.06$ fm for $^{8}$He \cite{Tani92}.
Hence, $a=1.40$ fm and $b=2.04$ fm for $^{6}$He, $a=1.43$ fm and
$b=1.93$ fm for $^{8}$He and $a=1.325$ fm for $^{4}$He. Thus, the
rms radii of the point-proton density distributions are 1.62 fm,
1.72 fm and 1.76 fm for $^{4}$He, $^{6}$He and $^{8}$He,
respectively.

We use secondly the COSMA point nucleon densities \cite{Zhuk94} of
$^{6,8}$He which have the same analytical form (\ref{eq:rhotani})
and are based on the assumption of the $1p_{3/2}$ state for the
relative motion of the alpha-core and each of the valence
neutrons. The parameter values are \cite{Korsh197,Korsh297}
$a=1.55$ fm, $b=2.24$ fm for $^{6}$He and $a=1.38$ fm, $b=1.99$ fm
for $^{8}$He. The rms radii of the point-proton density
distributions are 1.89 fm and 1.69 fm for $^{6}$He and $^{8}$He,
respectively. Here we would like to note the different trend in
the behavior of the rms radii of the point-proton density
distributions in the Tanihata density \cite{Tani92} and COSMA
density \cite{Zhuk94,Korsh197,Korsh297}. While in the former the
rms radius increases from $^{4}$He to $^{6}$He and $^{8}$He, in
the latter it decreases from $^{6}$He to $^{8}$He. Thus both
densities contain opposite effects of the additional neutrons on
the charge distributions in the helium isotopes. We note also that
both Tanihata and COSMA densities have a Gaussian asymptotic
behavior which is supposed not to be a realistic one for these
nuclei at high $q$. Both densities had been used to fit mainly the
total reaction cross sections and the rms radii. That is why it is
necessary to consider also other more realistic theoretical
predictions for the densities of the helium isotopes. We use also
the proton densities of $^{4,6,8}$He obtained within the LSSM
calculations in a complete 4$\hbar\omega$ shell-model space
\cite{Kara2000}. In them Woods-Saxon single-particle (s.p.) wave
functions have been used for $^{6}$He and $^{8}$He and
harmonic-oscillator ones for $^{4}$He. We use also the
experimental charge density for $^{4}$He \cite{DeVries87} with
charge rms radius equal to 1.696(14) fm.

\subsection{Density distributions of $^{6}$Li and $^{11}$Li}

The COSMA point nucleon densities used in our calculations for
$^{6}$Li and $^{11}$Li can be written in the following form
\cite{Korsh197}:
\begin{equation}
\rho_{point}^{X}(r)=N_{cX}\frac{\exp(-r^{2}/a^{2})}{\pi^{3/2}a^{3}}+
N_{\nu X}\frac{2\exp(-r^{2}/b^{2})}{3\pi^{3/2}b^{5}} \left[
Ar^{2}+\frac{B}{b^{2}}\left(r^{2}-\frac{3}{2}b^{2}\right
)^{2}\right ],\;\;X=Z,N.
\end{equation}
\label{eq:licosma}
For $^{6}$Li (in the $\alpha $+2$N$ model): $A$=1, $B$=0,
$N_{cZ}=N_{cN}=2$, $N_{\nu N}=N_{\nu Z}=1$, $a$=1.55 fm and
$b$=2.07 fm \cite{Korsh297}. For $^{11}$Li (in the $^{9}$Li+2$N$
model): $A$=0.81, $B$=0.19, $N_{cZ}=3$, $N_{cN}=6$, $N_{\nu Z}=0$,
$N_{\nu N}=2$, $a$=1.89 fm and $b$=3.68 fm \cite{Korsh197}. The
rms radii of protons are 2.44 fm and 2.31 fm for $^{6}$Li and
$^{11}$Li, respectively. Similarly to the case of the helium
isotopes, the COSMA density predicts a decrease of the proton rms
radius from $^{6}$Li to $^{11}$Li when the number of neutrons
increases. Secondly, we use the proton density of $^{6}$Li
obtained within a LSSM in a complete 4$\hbar\omega$ shell-model
space and of $^{11}$Li in a complete 2$\hbar\omega$ shell-model
calculations \cite{Kara97}. For $^{11}$Li Woods-Saxon s.p. wave
functions have been used and harmonic-oscillator ones for
$^{6}$Li. Third, for $^{6}$Li we use as well the density
distribution of point-like protons \cite{Patter2003,Fried94} which
leads to the experimental charge distribution with rms radius
equal to 2.57 fm \cite{Patter2003}.

\subsection{Density distributions of $^{17,19}$B and $^{14}$Be}

The point nucleon distributions in $^{17,19}$B and $^{14}$Be were
determined in \cite{Suzuki99} from analyses of the interaction
cross sections of light radioactive nuclei close to the neutron
drip line measured at around 800$A$ MeV using the optical limit to
Glauber type calculations by employing a harmonic-oscillator type
densities including the contribution from the $sd$-shell:
\begin{equation}
\rho_{point}^{Z}(r)=\frac{2}{\pi^{3/2}\lambda^{3}}\frac{1}{\left(1-\frac{1}{A}\right
)^{3/2}}\exp(-x^{2})\left [1+\frac{(Z-2)}{3}x^{2}\right ],
\label{eq:bbez}
\end{equation}
\begin{equation}
\rho_{point}^{N}(r)=\frac{4}{\pi^{3/2}\lambda^{3}}\frac{1}{\left(1-\frac{1}{A}\right
)^{3/2}}\frac{N}{(N+8)} \exp(-x^{2})\left [1+2x^{2}+
\frac{(N-8)}{15}x^{4}\right ],
\label{eq:bben}
\end{equation}
where
\begin{equation}
x^{2}=\left(\frac{r}{\lambda}\right )^{2}\frac{A}{A-1}
\label{eq:x2}
\end{equation}
and $\lambda $ is the width parameter. Its values are determined
by a fit to the experimental interaction cross sections
\cite{Suzuki99,Tani85}. In our work we use the values of $\lambda$
which correspond to the values of the effective rms radii of the
point nucleon distributions obtained in \cite{Suzuki99}. They are
1.9366 fm, 2.0365 fm and 2.0218 fm for $^{17}$B, $^{19}$B and
$^{14}$Be, respectively. The rms radii of the proton distributions
are correspondingly 2.722 fm, 2.872 fm and 2.755 fm for $^{17}$B,
$^{19}$B and $^{14}$Be. Similarly to the case of the Tanihata
densities of the helium isotopes the proton rms radius increases
from $^{17}$B to $^{19}$B.

\subsection{Results of calculations}
In this subsection we present and compare the charge form factors
$F_{ch}(q)$ for different nuclei calculated using Eq.
(\ref{eq:fffolded}) (which accounts for the proton size and in
which the positions of the protons are determined from the
centre-of-mass of the nucleons in the nucleus) and also make a
comparison in one case with a form factor for point-like protons.

\begin{figure}[th]
\centering\epsfig{file=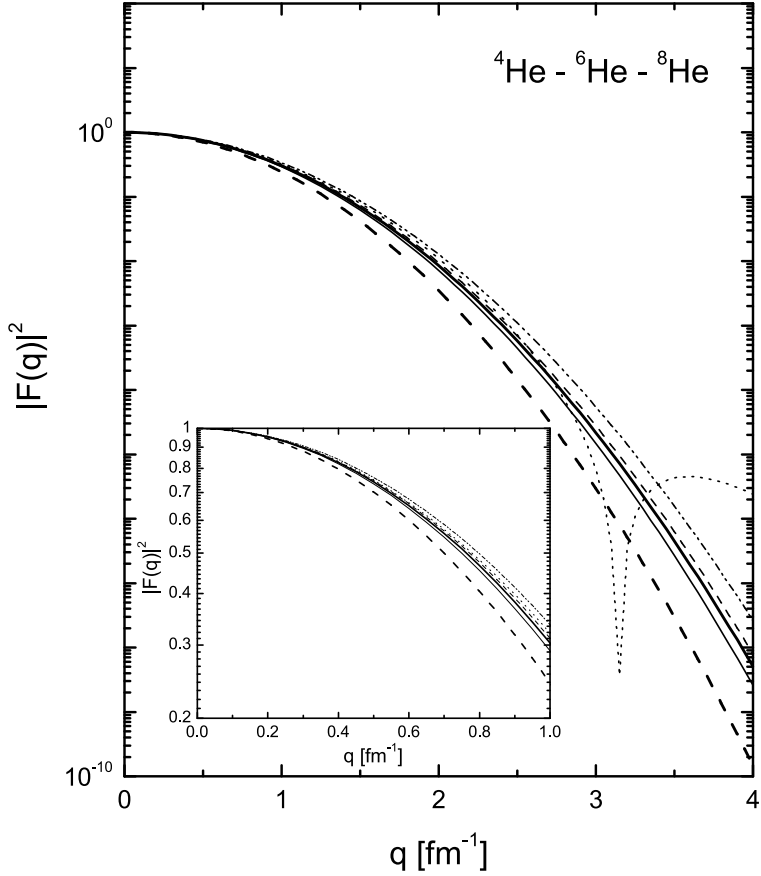,width=10cm} \caption[]{Charge form
factors from Eq. (\protect\ref{eq:fffolded}): a) for $^{4}$He
calculated with the experimental $\rho_{ch}(r)$
\protect\cite{DeVries87} (dotted line) and with the Tanihata
density \protect\cite{Tani92} (dash-two-dotted line); b) for
$^{6}$He calculated using the Tanihata \protect\cite{Tani92}
(thick solid line) and the COSMA \protect\cite{Korsh197} (thick
dashed line) densities, and c) for $^{8}$He calculated using the
Tanihata \protect\cite{Tani92} (thin solid line) and the COSMA
\protect\cite{Korsh197} (thin dashed line) densities.}
\label{fig1}
\end{figure}

\begin{figure}[th]
\centering\epsfig{file=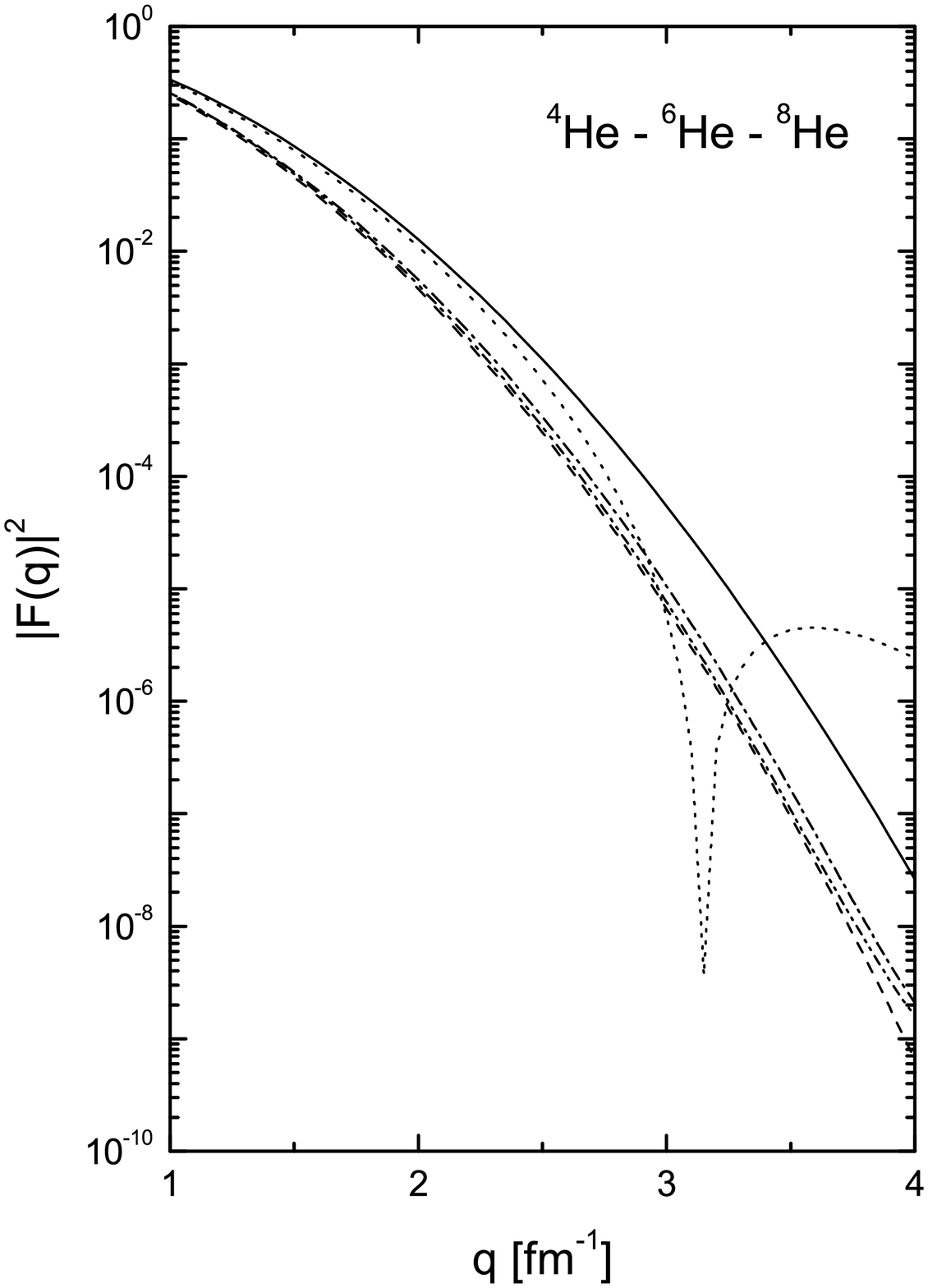,width=8cm} \caption[]{Charge form
factors for $q\geq $ 1 fm$^{-1}$ from Eq.
(\protect\ref{eq:fffolded}): a) for $^{4}$He calculated with the
experimental $\rho_{ch}(r)$ \protect\cite{DeVries87} (dotted
line), with the Tanihata density \protect\cite{Tani92} (solid
line) and using the LSSM density \protect\cite{Kara2000} (dashed
line); b) for $^{6}$He calculated using the LSSM density
\protect\cite{Kara2000} (dash-dotted line), and c) for $^{8}$He
calculated using the LSSM density \protect\cite{Kara2000}
(dash-two-dotted line).} \label{fig2}
\end{figure}

\begin{figure}[th]
\centering\epsfig{file=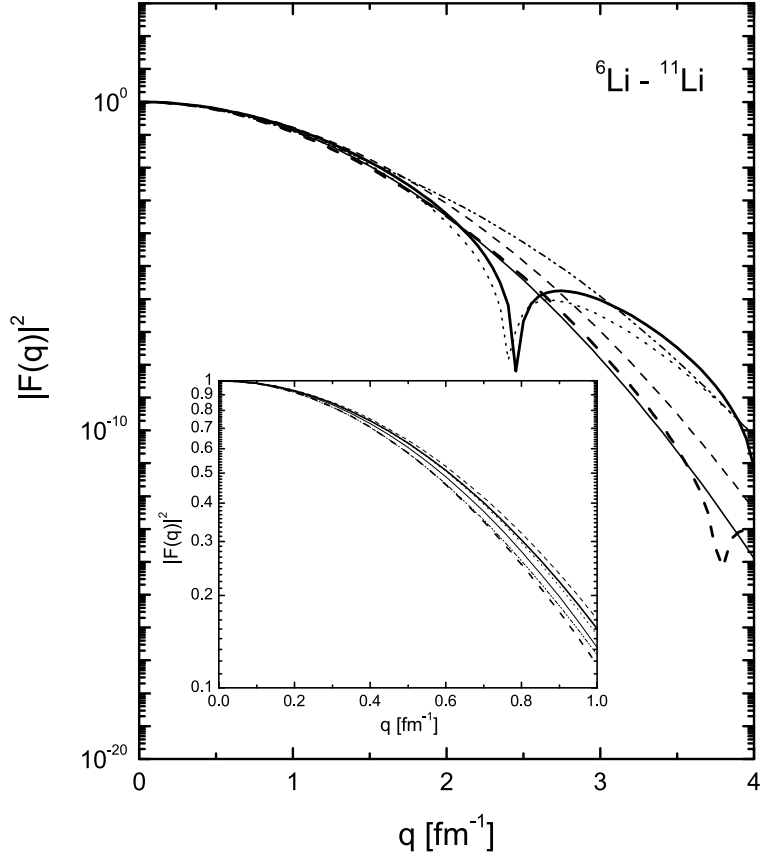,width=10cm} \caption[]{Charge form
factors from Eq. (\protect\ref{eq:fffolded}): a) for $^{6}$Li
calculated using the experimental $\rho_{ch}(r)$
\protect\cite{Patter2003,Fried94} (dotted line), using the COSMA
density \protect\cite{Korsh197,Korsh297} (dash-two-dotted line),
and using the LSSM density \protect\cite{Kara97} (thick dashed
line), and b) for $^{11}$Li calculated using the the COSMA density
\protect\cite{Korsh197} (thin solid line), the point-proton form
factor $F_{point}(q)$ calculated using the point-proton density
from \protect\cite{Korsh197} (thin dashed line), and using the
LSSM density \protect\cite{Kara97} (thick solid line).}
\label{fig3}
\end{figure}

\begin{figure}[th]
\centering\epsfig{file=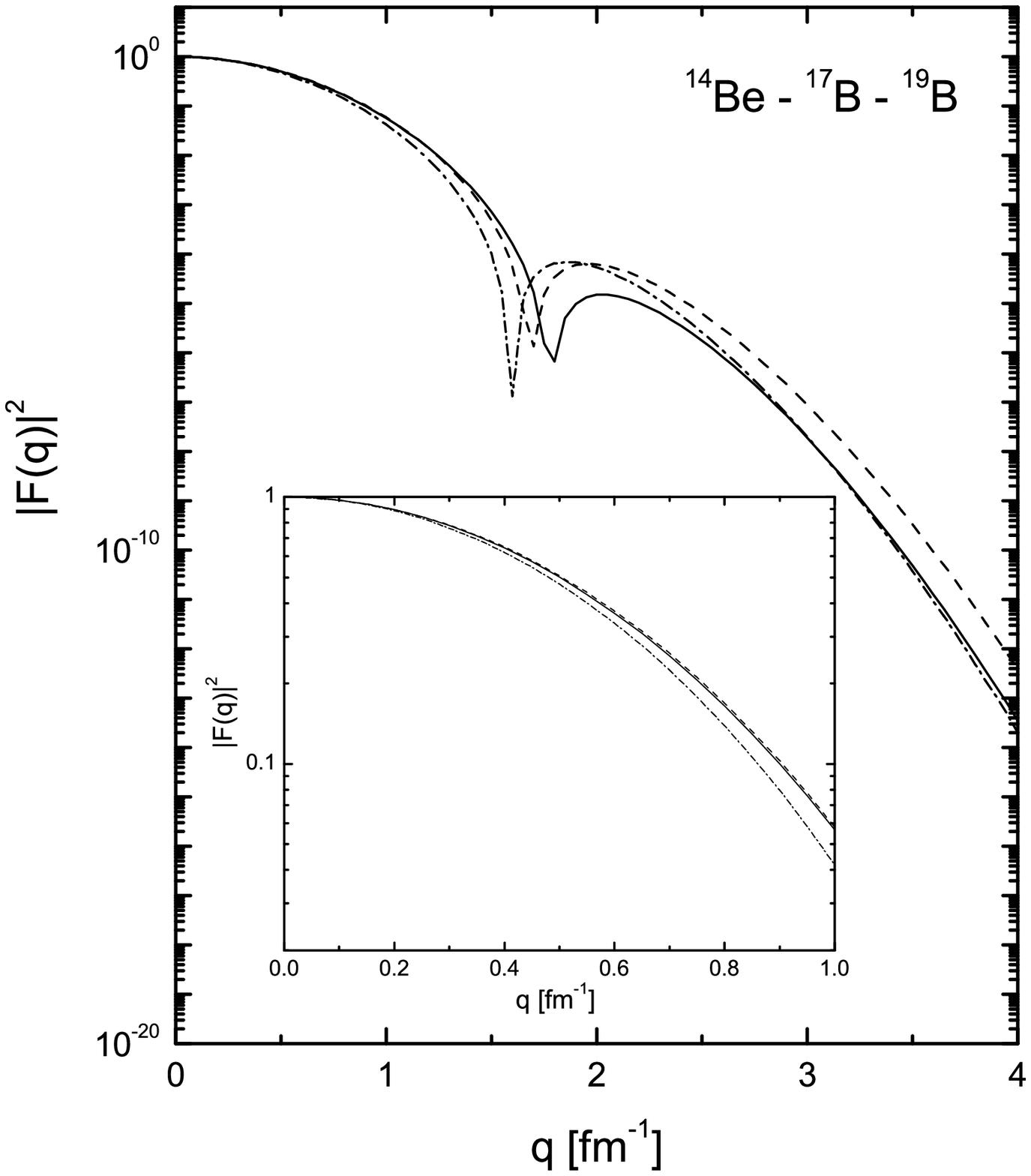,width=10cm} \caption[]{Charge form
factors from Eq. (\protect\ref{eq:fffolded}) for $^{14}$Be (solid
line), $^{17}$B (dashed line) and $^{19}$B (dash-dotted line)
calculated using the point-proton densities from
\protect\cite{Suzuki99}.} \label{fig4}
\end{figure}

In Fig. \ref{fig1} are compared the charge form factors
$F_{ch}(q)$ (\ref{eq:fffolded}) of $^{6}$He and $^{8}$He
calculated using the Tanihata \cite{Tani92} and the COSMA
\cite{Korsh197} densities. For comparison are given the charge
form factors for $^{4}$He calculated using the experimental charge
distribution from \cite{DeVries87} and the Tanihata density
\cite{Tani92}. In Fig. \ref{fig2} the charge form factors for
$^{4,6,8}$He calculated using the LSSM densities \cite{Kara2000}
are compared with those for $^{4}$He calculated with the
experimental $\rho_{ch}(r)$ \cite{DeVries87} and with the Tanihata
density \cite{Tani92}. In Fig. \ref{fig3} the charge form factors
from (\ref{eq:fffolded}) for $^{6}$Li and $^{11}$Li calculated
using the COSMA densities \cite{Korsh197,Korsh297} are given
together with those from calculations using the LSSM proton
densities \cite{Kara97}. For comparison are given the charge form
factors for $^{6}$Li calculated using the experimental charge
distribution from \cite{Patter2003,Fried94} as well as the
point-proton form factor for $^{11}$Li calculated with the COSMA
density distribution. The charge form factors of $^{17}$B and
$^{19}$B are compared in Fig. \ref{fig4} with that of $^{14}$Be
using Eq. (\ref{eq:fffolded}) and the point-proton densities from
\cite{Suzuki99}. The charge rms radii and the diffuseness
parameters of the charge densities are given in Table 1. The
diffuseness parameter is defined as the distance over which the
value of $\rho_{ch}(r)$ decreases from 90\% to 10\% of the value
of the density in the centre of the nucleus $\rho_{ch}(r=0)$
divided by 4.4.

\subsection{Discussion and conclusions}
The main features of the obtained charge form factors can be
summarized as follows:

Although the number of the protons within a given group of
isotopes is the same, the charge form factors are quite different
from each other. This concerns $^{4,6,8}$He nuclei (using the
Tanihata, the COSMA and the LSSM densities), $^{6}$Li and
$^{11}$Li (using the COSMA and the LSSM densities), as well as
$^{17}$B and $^{19}$B isotopes (using the Suzuki densities). These
differences reflect the differences in the phenomenologically
obtained proton density distributions (which have been fitted to
the empirical data for the total reaction cross sections) and in
the corresponding radii. Thus the charge form factors can show the
effects of the different number of neutrons and their distribution
in the nucleus on the charge distribution in a given group of
isotopes. Indeed, e.g. $^{4}$He is a strongly bound $\alpha
$-cluster, while $^{6}$He is a loosely bound $\alpha $+$n$+$n$
system. $^{6}$Li is predominantly an $\alpha $+$d$ formation while
$^{11}$Li is supposed to be a $^{9}$Li+$n$+$n$ system. We should
emphasize, however, the different trend of the behavior of the
charge form factors for $^{6}$He and $^{8}$He when using the
Tanihata, the COSMA and the LSSM densities. This follows, as can
be expected, the changes of the proton rms radii (i.e. the changes
of the neutron effect on the charge distributions) within a given
theoretical approach. While in the case of the Tanihata density
the proton rms increases from 1.72 fm for $^{6}$He to 1.76 fm for
$^{8}$He, in the COSMA and LSSM cases it decreases from 1.89 fm to
1.69 fm (in COSMA) and from 1.95 fm to 1.92 fm (in LSSM) for
$^{6}$He and $^{8}$He.

We should mention that the Tanihata, COSMA and Suzuki densities
used in our work have Gaussian asymptotic behavior which is
supposed not to be a realistic one for the nuclei considered at
high $q$. This imposes the usage of other more realistic
theoretical densities that may become available. We used the LSSM
ones for $^{4,6,8}$He and $^{6,11}$Li nuclei. For $^{6}$He,
$^{8}$He and $^{11}$Li Woods-Saxon s.p. wave functions have been
used in the LSSM calculations.

We note also the important effect of the proton size accounted for
in the calculations of the charge form factors.

In summary, various existing models of charge densities for light
exotic nuclei differ very much in their form factor predictions.
They even predict opposite effects of the neutron excess on the
charge distribution that could be tested by measuring charge form
factors. Therefore, it would be most desirable to compare the form
factors obtained here with the future experimental data on
electron-exotic nuclei scattering. Our numerical results for the
charge form factors are available upon demand.

\section{Acknowledgments}
One of the authors (A.N.A.) thanks the Royal Society of London and
the Bulgarian Academy of Sciences for support during his visit to
the University of Oxford. He is also grateful for warm hospitality
to the Faculty of Physics of the Complutense University of Madrid
and for support during his stay there to the State Secretariat of
Education and Universities of Spain (N/Ref. SAB2001 - 0030). Three
of the authors (A.N.A., M.K.G. and D.N.K.) are grateful to the
Bulgarian National Science Foundation for partial support under
the Contract No. $\Phi $--905.

\noindent {\bf Table 1:} Charge radii (in fm) and diffuseness (in
fm) of the charge densities.

\begin{center}
\begin{tabular}{cccccc}
\hline
Nuclei & & Charge rms & Diffuseness & Densities \\
\hline

$^{4}$He    & & 1.68 & 0.30 & Exp. \cite{DeVries87} \\
            & & 1.81 & 0.41 & Tanihata \cite{Tani92} \\
            & & 2.03 & 0.44 & LSSM \cite{Kara2000} \\
$^{6}$He    & & 1.89 & 0.42 & Tanihata \cite{Tani92} \\
            & & 2.06 & 0.45 & COSMA \cite{Korsh197} \\
            & & 2.08 & 0.49 & LSSM \cite{Kara2000} \\
$^{8}$He    & & 1.93 & 0.43 & Tanihata \cite{Tani92} \\
            & & 1.87 & 0.41 & COSMA \cite{Korsh197} \\
            & & 2.04 & 0.45 & LSSM \cite{Kara2000} \\
$^{6}$Li    & & 2.57 & 0.54 & Exp. \cite{Patter2003,Fried94} \\
            & & 2.57 & 0.52 & COSMA \cite{Korsh197} \\
            & & 2.63 & 0.55 & LSSM \cite{Kara97} \\
$^{11}$Li   & & 2.45 & 0.54 & COSMA \cite{Korsh197} \\
            & & 2.51 & 0.57 & LSSM \cite{Kara97} \\
$^{14}$Be   & & 2.87 & 0.64 & Suzuki \cite{Suzuki99} \\
$^{17}$B    & & 2.84 & 0.60 & Suzuki \cite{Suzuki99} \\
$^{19}$B    & & 2.98 & 0.63 & Suzuki \cite{Suzuki99} \\
\hline
\end{tabular}
\end{center}

\end{document}